\documentclass[]{IEEEtran}

\ifCLASSINFOpdf
 \else
  
\fi

\usepackage{amsthm}
\usepackage{amsmath}
\usepackage{graphicx}
\usepackage[compatible]{algpseudocode}
\usepackage{amssymb}
\usepackage{listings}
\usepackage{amsfonts}
\usepackage{algorithm}
\usepackage{algcompatible}
\usepackage{tabularx}
\usepackage{lipsum}

\usepackage[utf8]{inputenc}

\usepackage{url}
\usepackage{hyperref}
\usepackage{balance}
\usepackage{subcaption}
\usepackage{epstopdf}
\usepackage{array}
\usepackage{comment}
\usepackage[noadjust]{cite}
\usepackage{amsthm}
\usepackage{xcolor}
\usepackage{breakurl}

\def\Plus{\texttt{+}}
\begin{document}
\title{Covid-19 Spread Detection and Controlling with Fog-based Infection Probability Evaluation Model}
\author{Suraj~Mahawar, and  Ajay~Pratap,~\IEEEmembership{Member,~IEEE}
	\IEEEcompsocitemizethanks{\IEEEcompsocthanksitem S. Mahawar and A. Pratap are with the Department of Computer Science and Engineering, Indian Institute of Technology (Banaras Hindu University) Varanasi 221005 India. E-mail: \{suraj.mahawar.cse19, ajay.cse\}@iitbhu.ac.in.\protect 
    	       }
	}
\maketitle

\begin{abstract}
COVID-19 has created a pandemic around the world, paused the path of building the future, and still ongoing without having any long-term solution shortly. The time taken in vaccine distribution is too slow compared to the spread of COVID-19. Hence, it is important to aware and take precautions on time without delaying and waiting for long-duration after getting infected with the virus. Currently used technology is more advanced than ever before. Almost everyone has access to at least one mobile device with an Internet connection. Therefore, we propose a Fog Server (FS) based system that can be used to create awareness about the spread of COVID-19 within the surroundings of individuals utilizing the concept of Hidden Markov Models (HMM) and Bluetooth contact tracing, in polynomial computational time complexity. Moreover, we evaluate the effectiveness of the proposed model through real-world data analysis on different simulation parameter settings. 

\end{abstract}

\IEEEpeerreviewmaketitle

\section{Introduction}

The COVID-19 disease started to spread in 2019 and has spread to over 180 countries. Some countries are facing the second wave of the virus spread right now. As of the 18th May 2021, a total of 164,355,605  cases and 3,406,601 deaths were reported \cite{cases}. The virus spreads through respiratory droplets or air from one person to another. These droplets spread when an infected person sneezes, coughs, or speaks to one or many. Infection can also spread if a person touches a contaminated surface and then eyes, nose, or mouth. Most common symptoms of COVID-19 include fever, dry cough, tiredness etc. If the lungs get completely infected the chances of death become high.

The COVID-19 is an infectious disease caused by the coronavirus which is a group of Ribonucleic Acid (RNA) viruses. Many people, if gather at places tends to get infected with COVID-19 in a short time and can experience respiratory illness and possibly recover without the need for special treatment. Elderly people tend to have more impact on COVID-19 as they have weaker immune systems and most have underlying medical problems such as diabetes, cancer and some other respiratory-related diseases. These medical problems can develop more weakness and increase the sickness along with COVID-19. It is known that COVID-19 affects different people in different ways, most people who are infected will develop illness and possibly recover without any special medication treatment. It is observed that fever, dry cough, and tiredness are the most common symptoms among infected people and symptoms like pains, soreness throat, diarrhea, dizziness, headache, loss of taste and smell, rashes over the skin are less common among the infected people. Serious symptoms are such as difficulty in breathing, chest pain, loss of speech and few other symptoms can be noticed among the infected people. On average it takes 2-14 days for the symptoms to show after meeting the infected person \cite{time}. Therefore early identification of an infected person is very important to stop the further spread of the virus by imposing certain measures.

With the help of Fog Servers (FS), our system will make people aware of their health conditions and their surroundings \cite{yasser2020covid}. Those who live in crowded areas like urban cities, mostly have access to at least one mobile device with internet, so it is possible to set up an application into their device to track their movements and their contact with others using Bluetooth options. It can be efficient to design an algorithm that works with this technology to aware people from time to time. Thus, motivated by the above-mentioned scenarios, this paper evolves the technique to create awareness among people by using contact traced data and their data with an algorithm framework that can find accurate results and track COVID-19 spread. We have designed the spread of the COVID-19 virus phenomenon by applying Hidden Markov Model (HMM) and Bluetooth contact tracing framework. This statistical tool is used for modeling the generative sequences characterized by a set of observable sequences. However, HMM framework is used to model all stochastic processes in real time problems. In our proposed model, the non-observable state of the system is controlled by a Markov process and the observable sequences of the system have underlying probabilistic dependencies.

In the current situation of the COVID-19 pandemic, a solution is needed to prevent and control virus spread. In this regard, the proposed framework can be used to aware people of the COVID situation around them. In nutshell, the main contributions of the work are summarised in the following:
\begin{itemize}
    \item\textbf{FS to detect COVID-19 spread}: An FS framework is used with HMM and Bluetooth contact tracing enabled to track every individual user. Fog interacts with all users and serves them with the best precautions and guidance needed as per the user's conditions.
    
    \item\textbf{Detect initial infected users}: Exponential spread of virus needs a solution to stop the spread in early stages. A chain starts from one user but may last up to million users, which means breaking the chain initially will help control virus spread. The proposed solution helps detecting initial infected users and prevent them to meet others so that their contribution to virus spread can be reduced.
    
    \item\textbf{Evaluate individual's infection probability}: Evaluation of individual's infection probability as well as risk levels. Probability helps identifying the user's status if a user is allowed to get in touch with others or not. Risk levels help creating health reports of users containing the best precautions needed and all necessary guidance from health exports according to the user's conditions.
    
    \item\textbf{Future prediction}: Future prediction is done with current data based on \emph{meetups}, fixed thresholds of time of contact, Bluetooth signal strength and threshold infection probability, which helps to get prepared for future conditions in advance and to fight against the virus with pre-planned strategies.
\end{itemize}

 The rest of the paper is organized as follows: Section \ref{Sec2} reviews the relevant work. The system model and the problem definition are introduced in Sections \ref{SM}. The proposed solution and performance study are given in Sections \ref{PS} and \ref{PerStdy}, respectively. Finally, Section \ref{ConFu} offers conclusions and future research directions.

\section{Related Works}\label{Sec2}

Recently researchers at SMU have come up with a 'Lab on a Chip' test which can identify COVID-19 immune response faster than current antibody tests \cite{SMU}. When someone gets infected, his body produces some antibodies to fight the virus, in that case, if these antibodies are present, it means that the person is infected with virus. This test is done by applying a blood drop to a micro-fluid chip and once applied a filter embedded in that chip extracts plasma from the sample of blood. This chip is then placed in an electronic instrument to detect the presence of specific antibodies in plasma. While the Defense Advanced Research Projects Agency (DARPA) developed a microchip to detect COVID-19 \cite{Darpa}. This microchip is used to constantly check an individual's blood for virus and once it detects COVID-19, it will alert the patient and the patient is advised to go for a blood test to confirm if the virus is present in the blood. But, it is hard to implant chips in everyone's body. It can take much time to do such operations. Thus, it is necessary to aware people of their surroundings. People these days are carrying smartphones everywhere they go and they use the internet which can be an opportunity for us in utilizing the power of technology to aware of the COVID-19 situation. It is possible to add an application into millions of mobile phones than implanting microchips into million people. There are still many other kind of researches going on, but taking advantage of technology, we would like to use mobile phones with internet connection and Bluetooth contact tracing to identify whether a person is infected with COVID-19 or not. Our method utilizes an algorithmic approach and contact tracing mechanism. Bluetooth is used in contact tracing. FS interacts with all users from time to time and serves them with the best precautions and guidance needed as per the user's conditions.

HMM \cite{rabiner1986introduction} has been used in many fields and applications, such as speech recognition, visual speech recognition, to identify contamination in water networks \cite{han2019aquaeis}, predicting traffic conditions \cite{qi2014hidden}, analyzing nosocomial pathogens transmission \cite{cooper2004analysis}, predicting transmembrane helices \cite{krogh2001predicting}, analyzing evolutionary rates in molecular sequences \cite{felsenstein1996hidden}, etc. Contact tracing \cite{eames2003contact} can drastically improve the analysis of the COVID-19 virus spread. It can tell people if they come into contact with an infected person. Then the person can quickly take action so that spread of the virus to their surroundings slows down/stops. Contact tracing is very useful and is being used in controlling infectious diseases \cite{armbruster2007contact}, impact of delays on virus spread de-escalation strategies \cite{kretzschmar2020impact}, Bluetooth-based contact tracing \cite{hernandez2020evaluating}, person-to-person covid-19 tracing \cite{ojagh2021person}. In the case of COVID-19, we have a hidden state and an observable state in virus spread, thus, the use of HMM is a better approach in finding COVID-19 spread.

\section{System Model and Problem Definition}\label{SM}

To detect and control the spread of COVID-19, we use FSs which utilize algorithms for detecting and controlling virus spread as well as prediction of future conditions. We further consider use of contact tracing for monitoring each user, that helps in predicting the chances of a user being infected with COVID-19 so that necessary actions can be taken in advance like the government imposing guidelines such as lockdown, strict use of masks or early distribution of vaccines in those locations to prevent further spread of the virus.

\begin{figure}
	\centering
	\includegraphics[height=4.7cm, width=8.7cm]{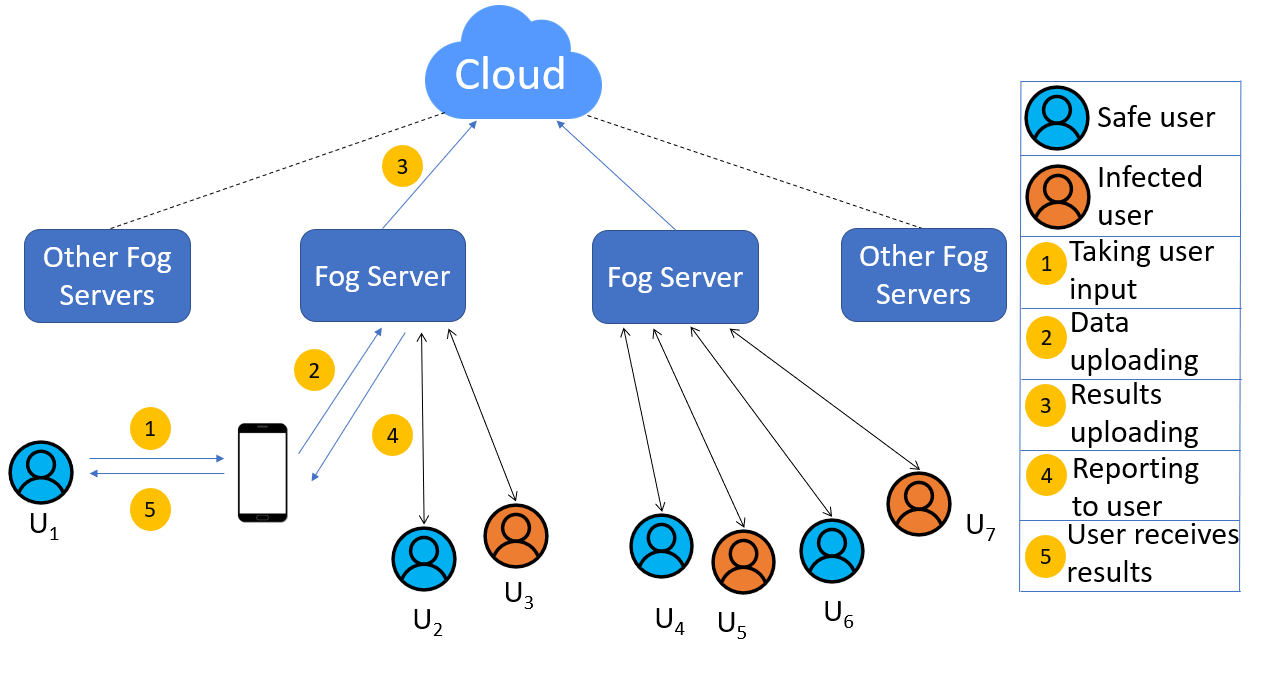}
	\vspace{-10pt}
	\caption{Fog based system model for COVID-19 spread identification.}
	\label{fig:fog}
\end{figure}

The system model represented in Fig. \ref{fig:fog} shows the mechanism of user data collection, user-given data uploading, data computation, storing the results and reporting, and further reporting the final results to the respective user and cloud server. The whole procedure is described in three sub-processes as follows:
\begin{itemize}
    \item Data collection and uploading to FS.
    \item Data computation at FS.
     \item Result analysis and reporting.  
\end{itemize}
Details of the above-mentioned sub-processes are discussed in the following sections:

 \subsubsection{Data collection and uploading to FS} 
 
 Our first aim is to collect user-given data. Initially, our system model registers a new user in the database with some identity as contact number, his current location, age, etc. Now user is asked to fill in some healthcare details, such as if the user has any symptoms related to COVID-19, if he had traveled to other infected countries/places within the last two weeks, and if user had any previous diseases. One more important input is the number of infected users, a user gets in touch with. It's obvious that a user can't predict whether the person near to him is infected with COVID-19 or not. In such a situation the system model needs some better way to find out the number of positive users, a person has got in touch with. Therefore, contact tracing is the best solution here to trace a person. Thus, we use Bluetooth connectivity for tracing other users near a particular user. When two persons come in the range of Bluetooth connectivity, it is called a \emph{meetup}. After a meetup, both user's devices connect and exchange some digital signatures. If any one of them is COVID-19 positive, our model sends an alert message to the user so that the user can get out of that particular location otherwise the user can also be infected. Thus, in this way, user given data and contact traced data are collected from users at the FS.
 
 \begin{figure}[h!]
	\centering
	\includegraphics[height=5cm, width=8.7cm]{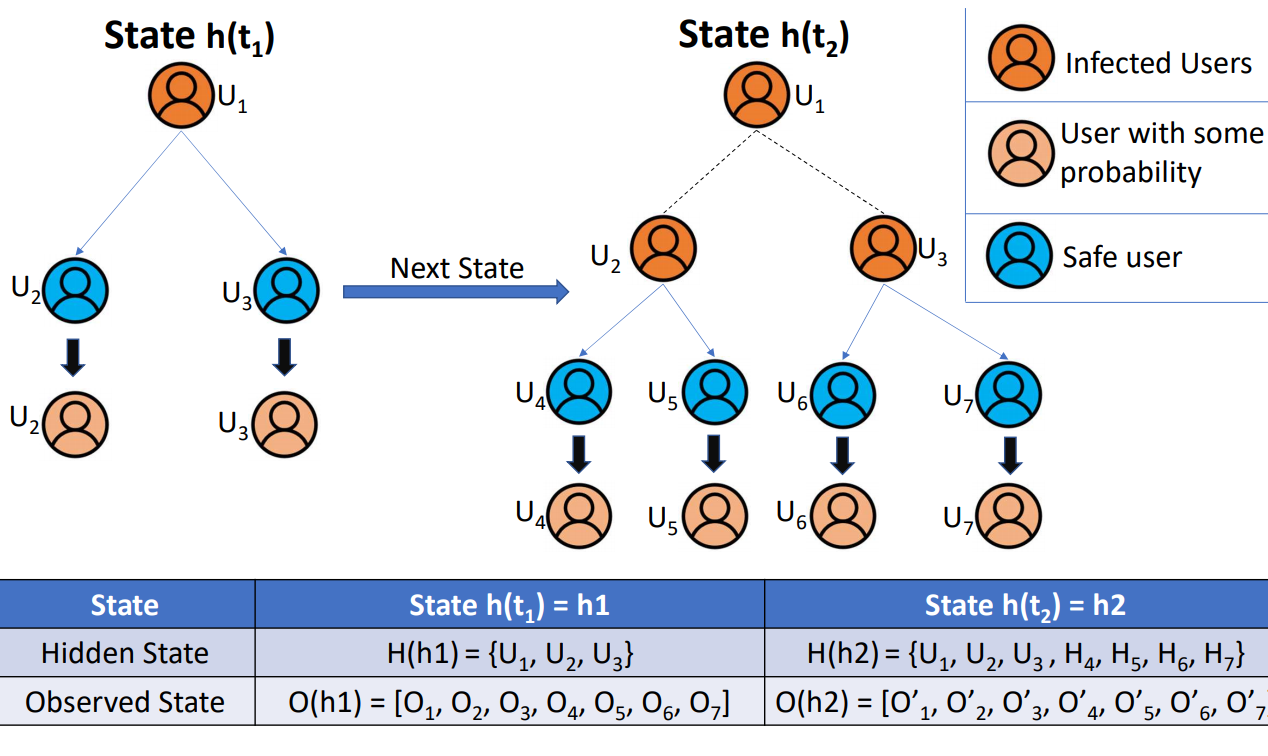}
	\caption{HMM model.}
	\label{HMM}
\end{figure}
 
\subsubsection{Data computation at FS}\label{DCFS} 

In this section, we compute the probability of infection by user-given data and contact traced data. The spreading of the virus can be modeled as an exponential problem as shown in Fig. \ref{HMM}, which shows two states $h(t_1)$ and $h(t_2)$. Here state means scenario of virus spread at a particular time slot. State $h(t_1)$ shows spreading of virus on time slot containing time $t_1$ and state $h(t_2)$ shows further spreading of virus on time slot containing time $t_2$. Fig. \ref{HMM} contains a table having hidden state and observed state named two states for each of $h(t_1)$ and $h(t_2)$ where $H$ is a set of infected users and $O$ is a set of each user's observations or set of all user given data and contact traced data. It is obvious that future conditions depend only on the current rate of virus spread and on the actions, we take today to control it. So, we have an exponential problem with some states and all the data (user-given data as well as contact tracing data) of those states. Now, we need to work on the current state's data and predict the most probable next state and also to analyze the results and provide the reports to users with all necessary precautions, medication and current conditions of particular user. Hence, we need to formulate algorithms, which work on user-given data as well as traced data to calculate probabilities of users being infected. Since the model works on previous states with all of their observed data (user given and traced data) and finds the most probable next state, so we formulate this future prediction as an HMM inference problem.

\subsubsection{Result analysis and reporting} 

After computing the user given and contact traced data at FS, system upload the results in a centralized cloud server for all users. After that, results are analyzed and reports are generated for each user and are sent to respective users. A report contains the risk level (in the form of low, moderate, high, very high) of a user being infected, all necessary precautions and medications based on user's conditions. The results are containing the probability of a user being infected which was computed at FS, contact traced data, user's previous diseases and current symptoms.

\section{Proposed Solution}\label{PS}

As discussed in the above subsection \ref{DCFS}, we use HMM modeling the problem. In order to achieve that, let there be $N$ users represented using a set $\mathbb{U}=\{U_1, \cdots, U_N\}$. Further, we define three sets as follows: The first set having $M$ users represented in a set of infected users $\mathbb{I}=\{I_1, \cdots, I_M\}$ which contains all the users having probability more than some threshold probability value $\Theta$ or have been tested positive and second one is a set of the traced number of users $\mathbb{T}=\{T_1, \cdots, T_N\}$ where $T_n$ means $n^{th}$ user $U_n$ has met $T_n$ number of users with some infection probability till time $t$ or in other words, user $U_n$ had $T_n$ number of \emph{meetups} with other infected users and the third one is set of symptom status $\mathbb{S}=\{S_1, \cdots, S_N\}$ where $S_n$ $\in$ $\{0,1\}$, which means if $n^{th}$ user $U_n$ is having any virus related serious symptoms, then $S_n = 1$ otherwise $S_n = 0$. All of these sets may change their data according to the changes in time/states.

The proposed algorithmic solution is divided into three parts as follows:
\begin{itemize}
    \item Probability calculation with contact traced data.
    \item Probability calculation with user's symptoms and evaluating total probability of infection.
    \item Result analysis and reporting to users and cloud.
\end{itemize}

\begin{figure}[h!]
	\centering
    \includegraphics[height=5cm, width=8.7cm]{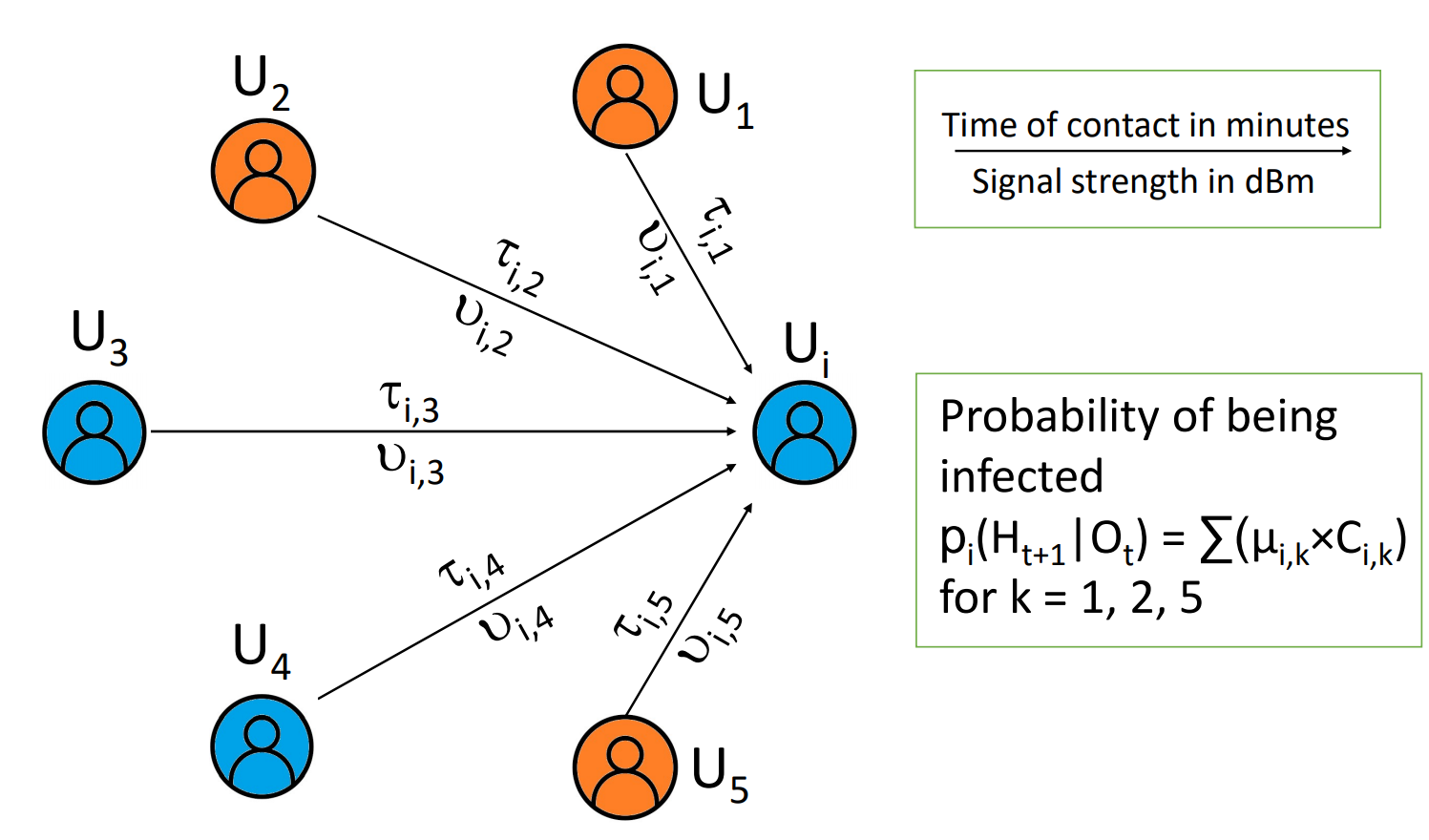}
	\caption{Probability calculations.}
	\label{sys}
\end{figure}

\subsubsection{Probability calculation with contact traced data.}

In this part, we aim to compute the probability of each user according to the previous state's results and current state's traced data. The basic idea and derivation of the equations are described with help of one example below.

\textbf{An illustrative example:} As given in Fig. \ref{sys}, there is a non-infected user $U_i$ and three infected users $U_1$,  $U_2$, $U_5$ with probability of more than some fixed threshold value $\Theta$. When user $U_i$ gets in touch with user $U_1$, then $U_i$ gains some probability, and if $U_i$ meets $U_2$, then also $U_i$ gets some more probability and so on. For further calculation, let probabilities of infection of users $U_1$, $U_2$, $U_5$ be $p_1$, $p_2$, $p_5$, respectively. Now, we aim to find that how much part of these probabilities should be added to the probability of user $U_i$. An optimal solution to this problem should be a variable that is dependent on several conditions like time for which two devices get connected ($\tau$), connection strength ($\nu$) (to get the space between two users based on signal strength), and some more measurements can also be used. For having an ideal solution and assuming ideal conditions, we use time of contact ($\tau$) and signal strength ($\nu$) as variables to define $\mu$, which would be called infection transition fraction and value of $\mu$ increases as the time of contact and signal strength increases. So, let $\mu_i(t) = \{ \mu_1, \mu_2, \mu_5 \}$, part of their probability is added to $U_i$, where $0 \leq \mu \leq 1$ which means that if user $U_i$ meets user $U_1$, then $U_i$ can only get maximum of $p_1$ probability from user $U_1$. Obviously, the probability can be at most 1, so if user meets so many infected users, then, at last, his probability gets 1 but it could also be greater than 1, so to prevent this condition, we use a boundary condition $0 \leq p_{i,t} \leq 1$, and if $p_{i,t} > 1$, then $p_{i,t} \gets 1$.

For using HMM, we need two states, the first one is the hidden state ($H_t$) and the second one is the observed state ($O_t$) at time $t$. The set of infected users, described above as $\mathbb{I}=\{I_1, \cdots, I_M\}$ is used as a hidden state because the model can't ensure whether a user is COVID-19 positive or not since our model can only predict some probability, as it can only be ensured after a lab test, that's why it can be called a hidden state. For the observed state, it should be directly observed by the model, as we can directly get the data by contact tracing then it is considered as an observed state. In our case, we use the set $\mathbb{T}=\{T_1, \cdots, T_N\}$ as observed state.

Let's calculate probability for $n^{th}$ user $U_n$ at time $t$. Let, at time $t$, hidden state is $H_t$ and observed state is $O_t$ and next future hidden state is $H_{t+1}$, then $p_n(H_{t+1}|O_t)$ represents next state's probability of infection of user $U_n$ due to meeting infected users of current state. Here we introduce a new set of sets of probabilities of traced infected users as $\mathbb{C} = \{C_1, \cdots C_N\}$ which contains sets of probabilities of traced infected users, where $C_n$ is set, containing all infected user's probability, who were traced within range of $n^{th}$ user and $C_{n,k}$ is $k^{th}$ user's probability of infection, who got in contact with $n^{th}$ user.

When user $U_n$ meets $T_n$ number of infected users, he gets some amount of each infected user's probability of infection. Let, $\mu$ distribution for $n^{th}$ user is $\mu_n = \{\mu_{n,1}, \cdots, \mu_{n,T_n}\}$. Thus, probability of user $U_n$ of being infected can be:
\begin{equation} \label{1st}
\begin{aligned}
    p_n(H_{t+1}|O_t) & = (\mu_{n,1} \times C_{n,1}) + (\mu_{n,2} \times C_{n,2})\\
    & + \cdots + (\mu_{n,T_n} \times C_{n,T_n})\\
    & = \sum_{k=1}^{T_n} \mu_{n,k} \times C_{n,k}
\end{aligned}
\end{equation}

Now, we aim to find values of $\mu_{n,k}$ $\forall$ $1$ $\leq$ $k$ $\leq$ $T_n$. As we have discussed the dependencies of $\mu$ over time of contact and signal strength, we'll calculate $\mu$ with help of those dependencies. Let, $\tau_0$ be threshold time for a user to be infected after being in touch with infected user, $\nu_0$ be the threshold signal strength (after certain experiments, we have $\mu_0 = -0.55$ $dBm$ Received Signal Strength Indicator (RSSI) value, which is calculated for two meter distance between two users), and if signal strength is more than threshold, then only a user can be infected by meeting an infected user. Let $\tau_{n,k}$ be the time for which users $U_n$ and $U_k$ get connected and $\nu_{n,k}$ is signal strength. If signal strength $\nu_{n,k}$ is less than threshold $\nu_0$, means they are at a distance of more than two meters, and user cannot be infected (in ideal cases, when there is not any other reason), and in that case value of $\mu_{n,k}$ will directly be assumed $0$, but if the signal strength is more than or equal to threshold then only we can say that in this case a user gets more infected as time of contact increases. So, in the case of more than threshold signal strength, infection spread to user is directly dependent on time of contact. Therefore we write:

\vspace{0.2cm}
$\mu_{n,k} = \{$
\vspace{0.1cm}

\hspace{1.3cm} $0$: $\nu_{n,k} < \nu_0$
\vspace{0.1cm}

\hspace{1.3cm} $\mu_{n,k} \propto \tau_{n,k}$: $\nu_{n,k} \geq \nu_0$
\vspace{0.1cm}

\hspace{1cm} $\}$
\vspace{0.1cm}

Now, for dealing with signal strength, we need a new variable, let $\lambda \in \{0, 1\}$, such that if the signal strength is less than the threshold, then $\lambda = 0$, otherwise, $\lambda = 1$. For the time of contact, where we have assumed that if time of contact is less than or equal to the threshold, then user gains some part of infected user's probability, otherwise, if time of contact exceeds threshold limit, then user gains whole part of infected user's probability. Here also, we need a new variable, let $\omega$, where $0 \leq \omega \leq 1$, which deals with the time of contact and provides a fraction value which is the fraction of infected user's probability. So,

\vspace{0.1cm}
$\omega_{n,k} = \{$
\vspace{0.1cm}

\hspace{1.3cm} $\frac{\tau_{n,k}}{\tau_0}$: $\tau_{n,k} \leq \tau_0$
\vspace{0.1cm}

\hspace{1.3cm} $1$: $\tau_{n,k} > \tau_0$
\vspace{0.1cm}

\hspace{1cm} $\}$
\vspace{0.1cm}

Here, $\omega$ is the fraction value of infection probability which needs to be added to not-infected user, when he meets an infected user. So, keeping in mind all of the above constraints, we come to a solution given below:
\begin{equation} \label{mu}
    \mu_{n,k} = \lambda_{n,k} \times \omega_{n,k}
\end{equation}
From the above Eqs. (\ref{1st}) and (\ref{mu}), we get:
\begin{equation} \label{p(t+1)withmu}
    p_n(H_{t+1}|O_t) = \sum_{k=1}^{T_n} \left ( \lambda_{n,k} \times \omega_{n,k} \times C_{n,k} \right ).
\end{equation}

Algorithm 1 calculates probability of infection which a user gets after getting in contact with infected users. As input, it takes contact traced data as time of contact, Bluetooth signal strength and probabilities of infected users. Here, probability is calculated for an individual user, based on their meetups with other users. For each user, first the signal strength is evaluated and if it passes the threshold, then time of contact is used to calculate probability. Finally, probability due to meeting infected users, for each user is calculated. In the worst case, time complexity for algorithm 1 would be O($N \times T_n$), where $N$ is the total number of user in the database and $T_n$ is the maximum number of meetups among all users.

\begin{algorithm}
    \caption{Probability calculation with traced data}
     \begin{algorithmic}[1]
         \State \textbf{Input:} Contact traced data $T_t$ or set of \emph{meetups} of each user, time of contact set $\tau_t$, signal strength set $\nu_t$, contact traced probabilities set $C_t$.
	     \State \textbf{Output:} Probability distribution $p(H_{t+1}|O_t)$, Eq. (\ref{p(t+1)withmu}).
    	 \State \textbf{Initialize:} Initializing Probability Distribution $p_n(H_{t+1}|O_t)$ = $0$ $\forall$ 0 $\leq$ $n$ $\leq$ $N$, $\mu_{n,k} = 0$ $\forall$ 1 $\leq$ $k$ $\leq$ $T_n$
    	\vspace{0.1 cm}
    	\State \textbf{for} $n$ = $1, \cdots, N$ \textbf{do}
    	\State \hspace{0.6 cm} \textbf{for} $k = 1, \cdots, T_{n,t}$
    	\State \hspace{1.2 cm} \textbf{if} $\nu_{n,k} \geq \nu_0$ \textbf{then}
    	\State \hspace{1.8 cm} $\lambda_{n,k} = 1$
    	\State \hspace{1.8 cm} \textbf{if} $\tau_{n,k} \leq \tau_0$ \textbf{then}
    	\State \hspace{2.4 cm} $\omega_{n,k} = \frac{\tau_{n,k}}{\tau_0}$
    	\State \hspace{1.8 cm} \textbf{else}
    	\State \hspace{2.4 cm} $\omega_{n,k} = 1$
    	\State \hspace{1.2 cm} \textbf{else}
    	\State \hspace{1.8 cm} $\lambda_{n,k} = 0$
    	\State \hspace{1.8 cm} $\omega_{n,k} = 0$
    	\State \hspace{1.2 cm} $\mu_{n,k} = \lambda_{n,k} \times \nu_{n,k}$
    	\State \hspace{1.2 cm} $p_n(H_{t+1}|O_t) = p_n(H_{t+1}|O_t) + \left (\mu_{n,k} \times C_{n,k} \right )$
    	\State \hspace{0.6 cm} \textbf{if} $p_n(H_{t+1}|O_t) > 1$ \textbf{then}
    	\State \hspace{1.2 cm} $p_n(H_{t+1}|O_t) = 1$
	\end{algorithmic}
\end{algorithm}

\vspace{0.4 cm}
\subsubsection{Probability calculation from user's symptoms and evaluating total probability of infection}

Since, it may be the case that without even being in touch with any other infected user, a user can be infected by the virus. When we talk about the COVID-19 virus, we shouldn't neglect that this virus can live for hours and days on some surfaces or objects. As it is obvious that after having some symptoms only, a user goes for a health checkup, and then he finds the results. So, it is also important to keep track of user symptoms for better results. For example: let us take two users $U_1$ and $U_2$ and both are having some symptoms of the virus, but user $U_1$ didn't meet any infected user means $T_1 = 0$ for user $U_1$ and user $U_2$ has met $T_2$ number of infected users.
In this case, obviously user $U_2$ has more probability of being infected but we can't neglect user $U_1$ also. So, we need a probability function that calculates the probability of infection a user gets due to his symptoms.

Further, we use a symptom status of each user as set $\mathbb{S}=\{S_1, \cdots, S_N\}$, which was described in above Section \ref{SM}. Its obvious that in situations like COVID-19, no one wants to get in touch with a user having any of the symptoms which are related to COVID-19 even if the user is really positive or not, and it would also be very helpful if we prevent users to meet those users who are having any COVID-19 related symptoms. So, if a user is having any symptoms then we are directly assuming the user infected and prevent other users to meet him. In other words, we are adding all users with symptoms in the set of infected users $\mathbb{I}$, because then only we can prevent other users to meet them.

Now, we aim to derive a probability equation such that it works on symptom status for each user and results in a probability that shows additional chances of infection due to symptoms. In the previous example of two users $U_1$ and $U_2$, both have serious symptoms but contact traced data for user $U_1$ is $T_1 = 0$ and for user $U_2$ is $T_2 > 0$. Now, we want to add both of them to the set of infected users, even if any one of them didn't meet a single infected user. Their current probabilities of being infected according to contact traced data are:

$(i)$ $U_1$: $p_1 = 0$, since he didn't meet any infected user.

$(ii)$ $U_2$: $p_2 = \sum_{k=1}^{T_2} \mu_{2,k} \times C_{2,k}$ for traced data $T_2$ and $C_2$.

Let the threshold probability for assuming a user infected is $\Theta$, then we need an equation which calculates probability in such a way that after adding that probability in $p_1$ and $p_2$, both of these users $U_1$ and $U_2$ gain more than $\Theta$ probability.

Let, probability due to symptoms is $\alpha_n$ and total probability of infection is $P_n$ (probability due to contact tracing ($p_{n,t}$) $\Plus$ probability due to symptoms ($\alpha_{n,t}$)) for $n^{th}$ user $U_n$ at time $t$.

$(i)$ For user $U_1$, contact tracing data as $T_1 = 0$, total probability of infection is:
$P_1 = p_1 + \alpha_1$;
where $P_1 \geq \Theta$ and $p_1 = 0$, so $\alpha_1 \geq \Theta$.

$(ii)$ For user $U_2$, contact tracing data as $T_2 > 0$, total probability of infection is:
$P_2 = p_2 + \alpha_2$;
where $P_2 \geq \Theta$ and $p_2 > 0$, so $\alpha_2 \geq \Theta - p_2$.

$(iii)$ For users having no symptoms as such: symptom status is $S = 0$ and probability due to symptoms is $\alpha = 0$.

So, at last we come to an optimal solution which is satisfying each of the above constraints as well as probability constraints, is derived below as:
\begin{equation} \label{symp_prob}
    \alpha_{n,t} = S_{n,t} \times (1 - p_{n,t}) \times \Theta
\end{equation}
where Eq. (\ref{symp_prob}) shows the probability of infection of $n^{th}$ user due to his symptoms. Hence, total probability can be calculated as:
\begin{equation} \label{tot}
    P_{n,t} = p_{n,t} + \alpha_{n,t}
\end{equation}
From above Eqs. (\ref{p(t+1)withmu}), (\ref{symp_prob}) and (\ref{tot}) we can write:
\begin{equation}\label{total_p}
\begin{aligned}
    P_{n,t} & = \left ( \sum_{i=1}^{T_n} \lambda_{n,i} \times \omega_{n,i} \times C_{n,i} \right ) \Plus \\
      & \left ( S_{n,t} \times \left ( 1 - \left  ( \sum_{i=1}^{T_n} \lambda_{n,i} \times \omega_{n,i} \times C_{n,i} \right ) \right ) \times \Theta \right ).
\end{aligned}
\end{equation}

We use number of newly infected users of each day as a result to compare model's work with real world data. In the following Algorithm 2, we evaluate newly infected users of each day, where $f_d$ shows number of newly infected users on day $d$. 
\begin{algorithm}
    \begin{algorithmic}[1]
        \State \textbf{Input:} probability distribution $p_{n,t}$ (from algorithm 1), symptom status $\mathbb{S}$, threshold probability $\Theta$
	    \State \textbf{Output:} probability due to symptoms $\alpha_{n,t}$, total probability $P_{n,t}$, $\forall$ $n \in \{1,\cdots,N\}$, time $t$ and number of infected users $f_d$ on day $d$
    	 \State \textbf{Initialize:} initialize $\alpha_{n,t} = 0$ $\forall$ $n$ $\in$ $\{1, \cdots, N\}$, $f_d = 0$
    	\vspace{0.1 cm}
	    \State \textbf{for} $n$ = $1, \cdots, N$ \textbf{do}
	    \State \hspace{0.6 cm} \textbf{if} symptoms are there \textbf{then}
	    \State \hspace{1.2 cm} $S_{n,t} = 1$
	    \State \hspace{0.6 cm} \textbf{else} 
	    \State \hspace{1.2 cm} $S_{n,t} = 0$
	    \State \hspace{0.6 cm} $\alpha_{n,t}$ = $S_{n,t} \times (1 - p_{n,t}) \times \Theta$
	    \State \hspace{0.6 cm} $P_{n,t} = p_{n,t} + \alpha_{n,t}$
	    \State \hspace{0.6 cm} \textbf{if} $P_{n,t} \geq \Theta$ \textbf{then}
	    \State \hspace{1.2 cm} $f_d = f_d + 1$
	\end{algorithmic}
	\caption{Probability calculation due to symptoms and evaluating total probability of infection} 
\end{algorithm}

In algorithm 2, probability due to the user's symptoms is calculated. First, symptom distribution is calculated, then with the proposed Eq. (\ref{symp_prob}), symptoms probability is calculated and total probability via adding both probabilities (probability due to traced infected users and probability due to symptoms) is calculated. After that, the number of newly infected users (total probability more than $\Theta$) is evaluated, and based on these results, graphs are plotted. Time complexity for Algorithm 2 with total of $N$ users in the database is O($N$).

\vspace{0.4 cm}
\subsubsection{Result analysis and reporting} Now, aim is to analyze the results, create reports for each user and send the reports to respective users. Results are also sent to the cloud to keep track of each user's health history and based on these results, reports are generated for individual user and reports are sent to them. These reports contain infection level of the respective users, precautions, medication and all necessary guidelines based on user's infection level, which help preparing the users to fight against the virus. The infection levels are initially divided into 4 parts: low, moderate, high, very high. Obviously, a low infection level means very few chances of being infected, moderate infection level means average chances of infection, high infection level means high chances of infection, and very high infection level means a user is infected. Moreover, the risk levels of all users are computed as follows:

$Infection$ $level$  = $\{$
\vspace{0.1cm}

\hspace{3cm} "Low" : $0 \leq P_{n,t} < P_l$
\vspace{0.1cm}

\hspace{3cm} "Moderate" : $P_l \leq P_{n,t} < P_{avg}$
\vspace{0.1cm}

\hspace{3cm} "High" : $P_{avg} \leq P_{n,t} < P_h$
\vspace{0.1cm}

\hspace{3cm} "Very High" : $P_{h}$ $\leq$ $P_{n,t}$ $\leq$ 1 
\vspace{0.1cm}

\hspace{3cm} $\}$

where thresholds for low is $P_l$, for moderate is $P_{avg}$, for high is $P_h$ and for very high is $1$, which helps creating a report based on user's current risk levels.

\begin{algorithm}
    \begin{algorithmic}[1]
        \State \textbf{Input:} Probability $P_t$ of all users being infected at time $t$, $P_l$, $P_{avg}$, $P_h$ thresholds
	    \State \textbf{Output:} Infection level $L_t = \{L_{1,t}, \cdots, L_{N,t}\}$ at time $t$
    	\State \textbf{Initialize:} initialize $L_t = [ \phi]$
    	\vspace{0.1 cm}
	    \State \textbf{for} $n$ = $1, \cdots, N$ \textbf{do}
	    \State \hspace{0.6 cm} \textbf{if} $0$ $\leq$ $P_{n,t}$ $<$ $P_l$ \textbf{then}
	    \State \hspace{1.2 cm} $L_{n,t}$ = "low"
	    \State \hspace{0.6 cm} \textbf{if} $P_l$ $\leq$ $P_{n,t}$ $<$ $P_{avg}$ \textbf{then}
	    \State \hspace{1.2 cm} $L_{n,t}$ = "moderate"
	    \State \hspace{0.6 cm} \textbf{if} $P_{avg}$ $\leq$ $P_{n,t}$ $<$ $P_h$ \textbf{then}
	    \State \hspace{1.2 cm} $L_{n,t}$ = "high"
	    \State \hspace{0.6 cm} \textbf{else}
	    \State \hspace{1.2 cm} $L_{n,t}$ = "very high"
	\end{algorithmic}
	\caption{Infection level calculation} 
\end{algorithm}

Let, at time $t$, $L_t$ be a set of $N$ elements, having values "low", "moderate", "high" or "very high", which means that user $U_n$ is having $L_{n,t}$ chances of being infected at time $t$. This result is provided to each user to make them ensure whether or not they are infected with virus or what are the chances or risks of infection. Separating users according to their risk levels is important because in this case a user is directly provided a report which tells him his current condition and the ways to be safe in the form of precautions, medications and many more healthcare benefits and help a user to make him aware of his condition.

Finally, FS has a set of infection levels, which helps FS creating reports for each user in an efficient way according to their infection level.

In Algorithm 3, aim is to get each user's risk levels based on their probability of infection. Each user is categorized in one of the four risk levels, and according to the user's risk level, reports containing precautions, healthcare medications, needed guidelines are generated and sent to respective users. Time complexity for Algorithm 3 is O($N$).

If we talk about overall time complexity of our proposed scheme, then in worst case, it would be O($N \times T_n$), where $N$ is the total number of users in the database and $T_n$ is the maximum number of meetups among all users, as described before.

\section{Performance Study}\label{PerStdy}

In this section, we illustrate the proposed scheme with help of different simulation parameters on real-time data. Here, we use two place's data of virus spread of 15 days. So, first, we evaluate those place's results with our model and then we compare them with real-world data. The data we are going to use are of two places, one is Pinal, Arizona, USA and the second one is Maricopa, Arizona, USA \cite{pinalmaricopa}.

We have taken 15 days of data of virus spread. For Pinal, the data is from the 20th March 2020 to 4th April 2020, and for Maricopa, it is from 15th March 2020 to 30th March 2020. The data shows the number of new COVID-19 confirmed cases on each day and are shown below in Fig. \ref{fig:Pinal_Maricopa} for Pinal and Maricopa.

\subsection{Model Setup}
In the simulation environment, 10,000 users are considered. Out of them, some may be suffering from the COVID-19 virus. On the first day, there will be a number of \emph{meetups}, and after each meetup between two users, if any one of them is infected, then the second user will gain some amount of infected user's infection probability according to the signal strength and time of contact which was described in Algorithms 1. After that, if the second user's probability of infection gets more than the fixed threshold ($\Theta$), then that user will also be assumed infected for future evaluations. Likewise, our model will evaluate for each day, for each meetup, and for each user. One more thing that is very important, is the distribution of the symptoms of users. As described in Algorithm 2, if someone gets any kind of COVID-19 related symptoms, probability due to that particular symptom is evaluated and added in the probability due to contact tracing. And finally, based on the results, reports are generated for each user with the help of Algorithm 3. So, this is how the infection spreads between several users and our model calculates each user's probability so that a user can be informed on early stages, which will help a user to take necessary precautions in advance and to fight easily against COVID-19.

\begin{figure}[h!]
\centering
	\includegraphics[height=4.5cm, width=8.7cm]{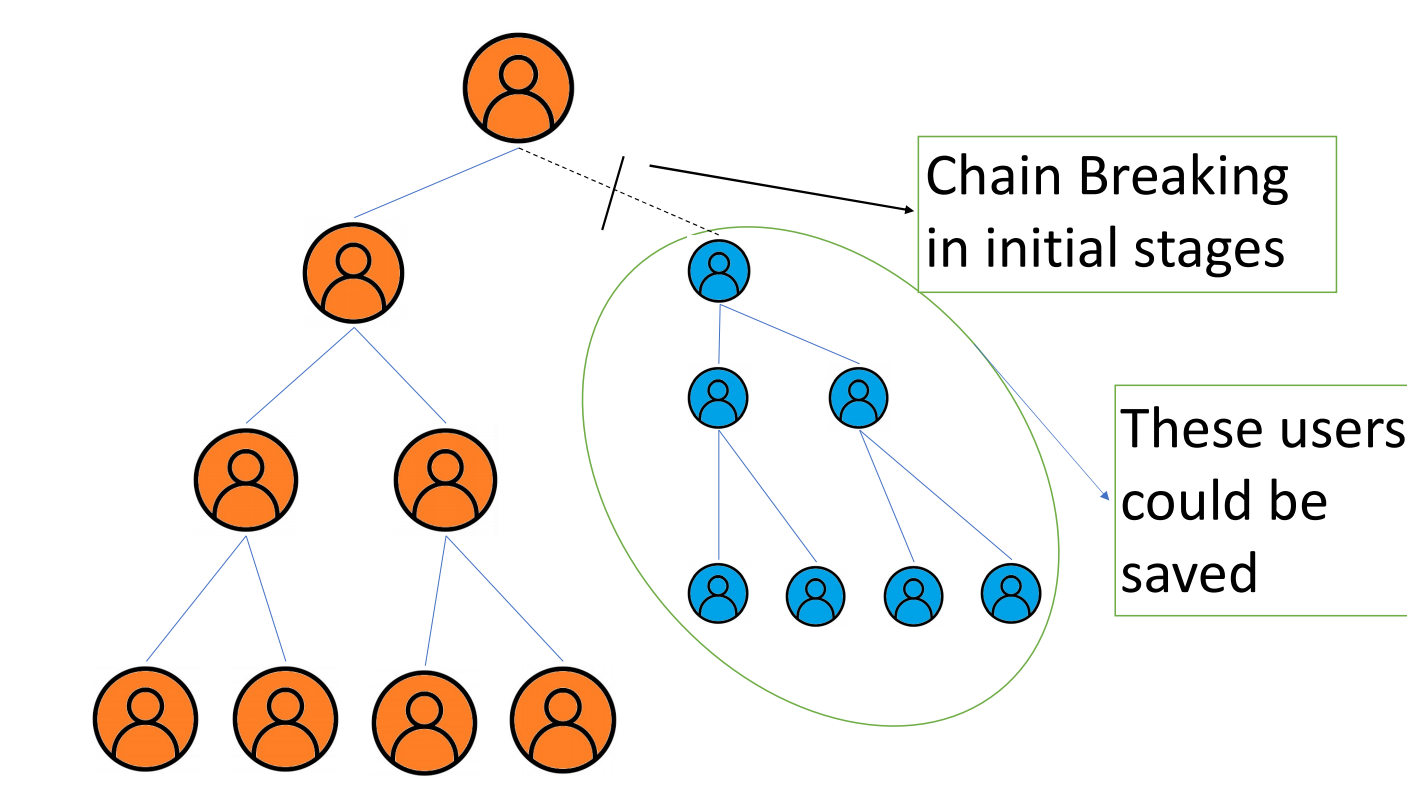} 
	\caption{Infection chain}
	\label{fig:chain_break}
\end{figure}

One most dangerous situation that is prevented by our model is that it helps in breaking the chain. For example, as given in Fig. \ref{fig:chain_break}, if a single user can be prevented from being infected in the early stages, then users who got infected by that user, can be saved and this can help to control the spread of the virus. So, our model alerts each user if he is going to meet an infected user which helps in breaking the chain and to control virus spread on the initial stage itself.

\subsection{Evaluation of Results}

As given in Algorithm 1, first probability of infection due to traced infected users is calculated then according to Algorithm 2, if symptoms are there for a particular user then probability due to those symptoms is added to that user and total probability is calculated. After applying both Algorithm 1 and Algorithm 2 on data of Pinal and Maricopa, the evaluated and real results are discussed below:

\begin{figure}[h!]
\centering
	\includegraphics[height=4.5cm, width=9cm]{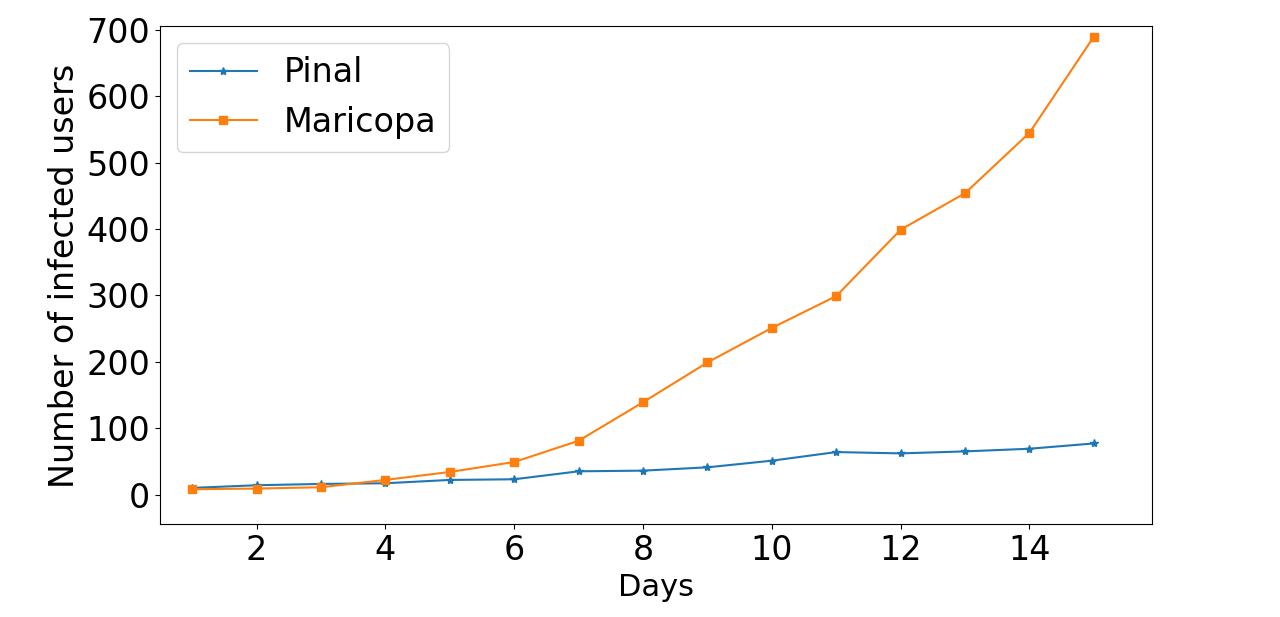}	\caption{Real time results of virus spread at Pinal and Maricopa.}
	\label{fig:Pinal_Maricopa}
\end{figure}

Fig. \ref{fig:Pinal_Maricopa} show the real-world data in form of a graph showing each day's number of newly infected users in Pinal and Maricopa. For example, in Pinal, on the first day (20th March 2020), the number of newly infected patients is 10, on the next day 14, and so on and in Maricopa, on the first day (15th March 2020), the number of newly infected patients are 8, on next day 9, and so on. Since, virus-infected users normally show symptoms after a long time of around 1 week to 2 weeks, which means that the results which are shown by graph are of 1 to 2 weeks earlier, means users had been infected 1 or 2 weeks before but it took some time for the virus to be exposed. In that case, if a user would be using the proposed model, then just after having some probability of infection, that user could be informed so that the user can easily fight against COVID-19 in earlier stages only.

Now, in the evaluation of results with our model, we use the term meetups, which was described above. Obviously, we can't know how many times a user will meet another user, in that case, we assume different cases but we can't only be dependent on just one case. So, here we work on 8 cases, from that 4 cases are for Pinal and 4 cases are for Maricopa. We use the different number of average meetups for each case and then finally for each case, graphs are plotted and comparisons are done. Moreover, the simulation parameters are shown in Table \ref{sim_param}. In Table \ref{sim_param}, rows 1-4 represents the simulation setup for the Pinal data set whereas the remaining rows are set up for the Maricopa data set for finding out the simulation results.

\begin{table}[h]
\caption{Simulation Parameters}
\begin{tabular}{|m{1cm}|m{1cm}|m{1cm}|m{1.2cm}|m{1.8cm}|}
    \hline
    Cases & theta $\Theta$ & $\tau_0$ (minutes) & $\nu_0$ (dBm) & Meetups a day\\
    \hline\hline
    1 & 0.9 & 2 & -0.55 & 1225 and 1250\\
    \hline
    2 & 0.9 & 2 & -0.55 & 1275 and 1300\\
    \hline
    3 & 0.9 & 1 & -0.50 & 1225 and 1250\\
    \hline
    4 & 0.9 & 1 & -0.50 & 1275 and 1300\\
    \hline
    5 & 0.9 & 2 & -0.55 & 2300 and 2330\\
    \hline
    6 & 0.9 & 2 & -0.55 & 2345 and 2360\\
    \hline
    7 & 0.9 & 1 & -0.50 & 2300 and 2330\\
    \hline
    8 & 0.9 & 1 & -0.50 & 2345 and 2500\\
    \hline
\end{tabular}\label{sim_param}
\end{table}

Here a question arises about the number of meetups like why only these number of meetups are assumed and why there is a difference between them. In answer to this question, we say that without deploying the system between people, we can't know which users are going to meet which and how many other users. So, here we have assumed that on an average day, there is an average of 1225 to 1300 meetups a day in Pinal and 2300 to 2360 meetups in Maricopa. The main reason behind the selection of these numbers is that these parameters are providing very good and satisfying results. If we talk about the difference between these meetups, then the reason is the population of both Pinal and Maricopa. With a population of 0.49 million in Pinal, it has less number of meetups compared to Maricopa with a 4.5 million population as it's obvious that highly populated areas will have a high number of meetups.

With all the parameters in the above simulation Table \ref{sim_param} and for each case, we evaluate results which are in form of several infected users of all days, after that a graph is plotted for each day's number of newly infected users. Each case having its separate parameters and data as given in the above table. So, evaluated results for each case are shown below in graphs.

\begin{figure}[h!]
\centering
	\includegraphics[height=4.5cm, width=8cm]{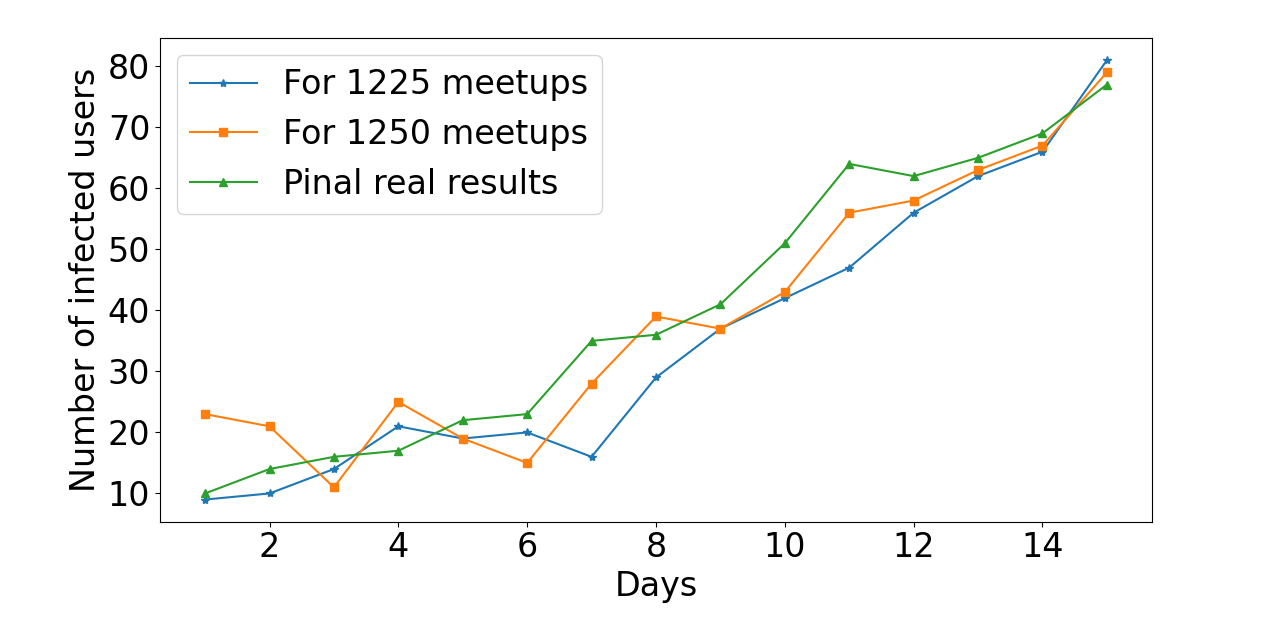}
	\caption{Case-1}
	\label{fig:case1}
\end{figure}

\textbf{Case-I:} In case-I, we work on Pinal dataset with threshold probability of infection $\Theta = 0.9$, threshold time of contact $\tau_0 = 2$ minutes, threshold signal strength $\nu_0 = -0.55$ dBm and assuming number of meetups as 1225 and 1250. As given in Fig. \ref{fig:case1}, three curves are plotted, first is for 1225 meetups, the second is for 1250 meetups and the third is for real-world results of Pinal.
 
\begin{figure}[h!]
\centering
	\includegraphics[height=4.5cm, width=8cm]{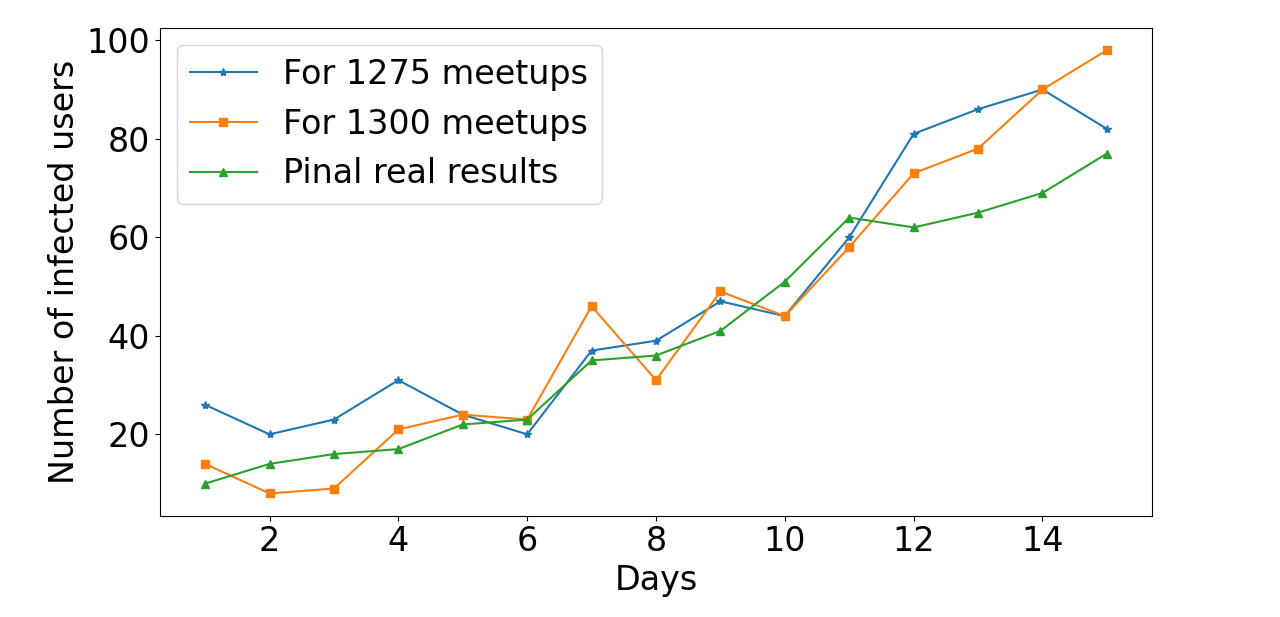}
	\caption{Case-2}
	\label{fig:case2}
\end{figure}

\textbf{Case-II:} In this case, we work on Pinal dataset with threshold probability of infection $\Theta = 0.9$, threshold time of contact $\tau_0 = 2$ minutes, threshold signal strength $\nu_0 = -0.55$ dBm and assuming number of meetups as 1275 and 1300. As given in Fig. \ref{fig:case2}, three curves are plotted, first is for 1275 meetups, the second is for 1300 meetups and the third is for real-world results of Pinal.

\begin{figure}[h!]
\centering
	\includegraphics[height=4.5cm, width=8cm]{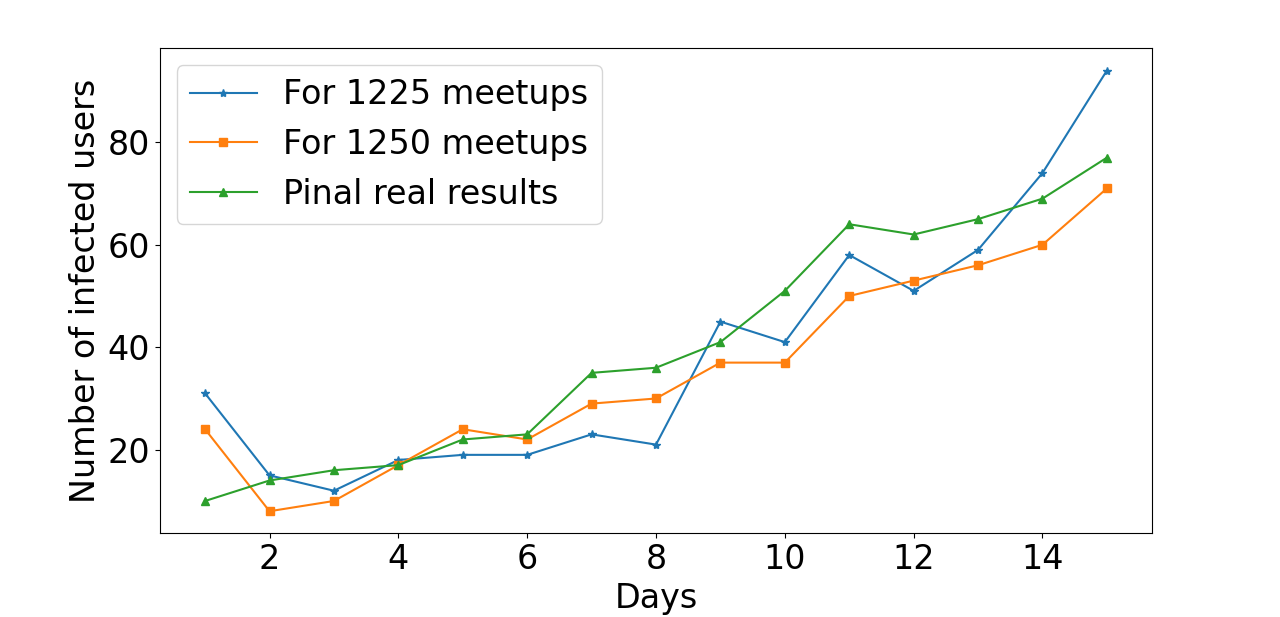}
	\caption{Case-3}
	\label{fig:case3}
\end{figure}

\textbf{Case-III:} In this case, we work on Pinal dataset with threshold probability of infection $\Theta = 0.9$, threshold time of contact $\tau_0 = 1$ minutes, threshold signal strength $\nu_0 = -0.50$ dBm and assuming number of meetups as 1225 and 1250. As given in Fig. \ref{fig:case3}, three curves are plotted, first is for 1225 meetups, the second is for 1250 meetups and the third is for real-world results of Pinal.

\begin{figure}[h!]
\centering
	\includegraphics[height=4.5cm, width=8cm]{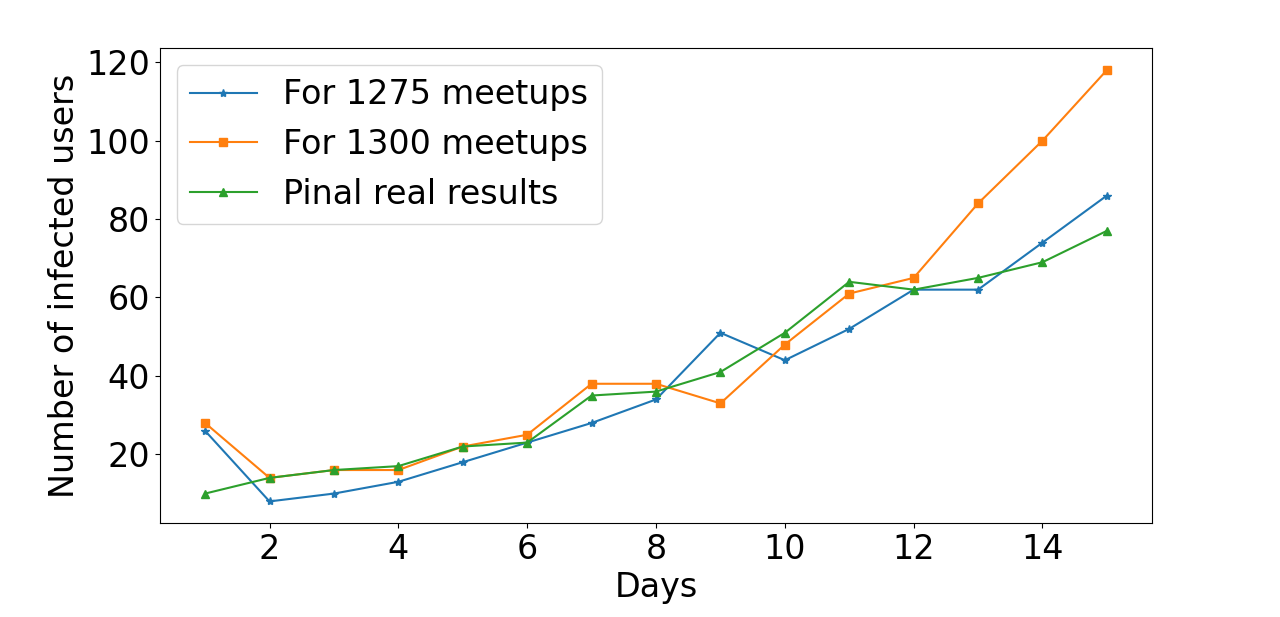}
	\caption{Case-4}
	\label{fig:case4}
\end{figure}

\textbf{Case-IV:} In this case, we work on Pinal dataset with threshold probability of infection $\Theta = 0.9$, threshold time of contact $\tau_0 = 1$ minutes, threshold signal strength $\nu_0 = -0.50$ dBm and assuming number of meetups as 1275 and 1300. As given in Fig. \ref{fig:case4}, three curves are plotted, first is for 1275 meetups, the second is for 1300 meetups and the third is for real-world results of Pinal.

\begin{figure}[h!]
\centering
	\includegraphics[height=4.5cm, width=8cm]{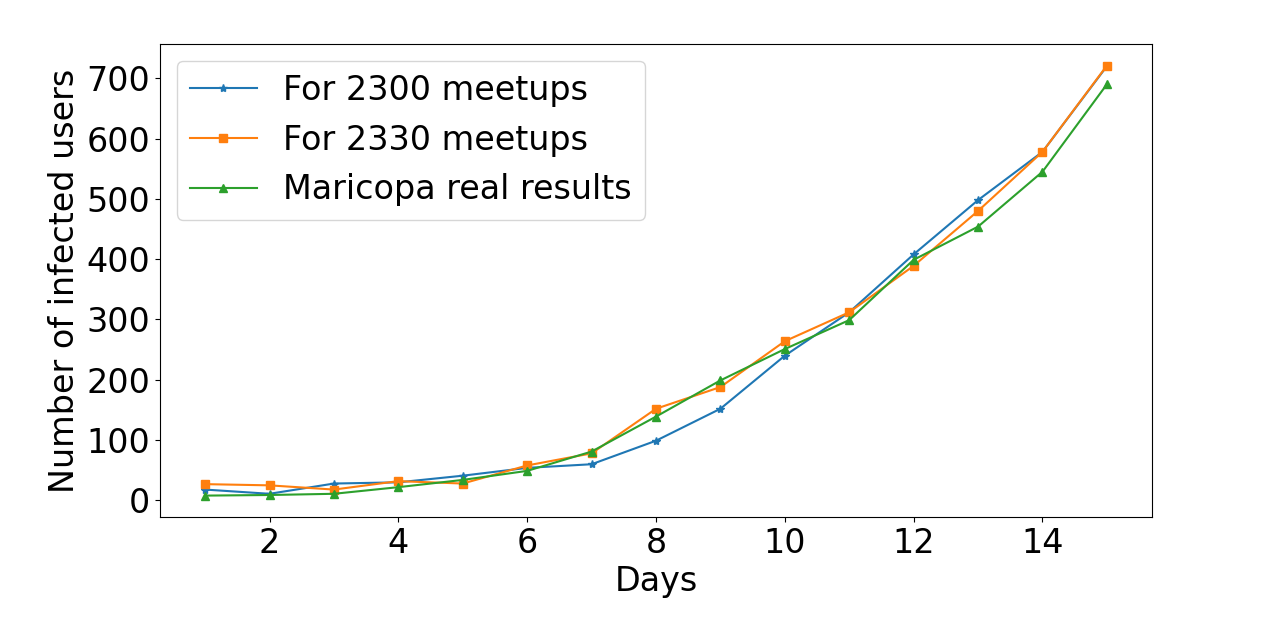}
	\caption{Case-5}
	\label{fig:case5}
\end{figure}

\textbf{Case-V:} In this case, we work on Maricopa dataset with threshold probability of infection $\Theta = 0.9$, threshold time of contact $\tau_0 = 2$ minutes, threshold signal strength $\nu_0 = -0.55$ dBm and assuming number of meetups as 2300 and 2330. As given in Fig. \ref{fig:case5}, three curves are plotted, first is for 2300 meetups, the second is for 2330 meetups and the third is for real-world results of Maricopa.

\begin{figure}[h!]
\centering
	\includegraphics[height=4.5cm, width=8cm]{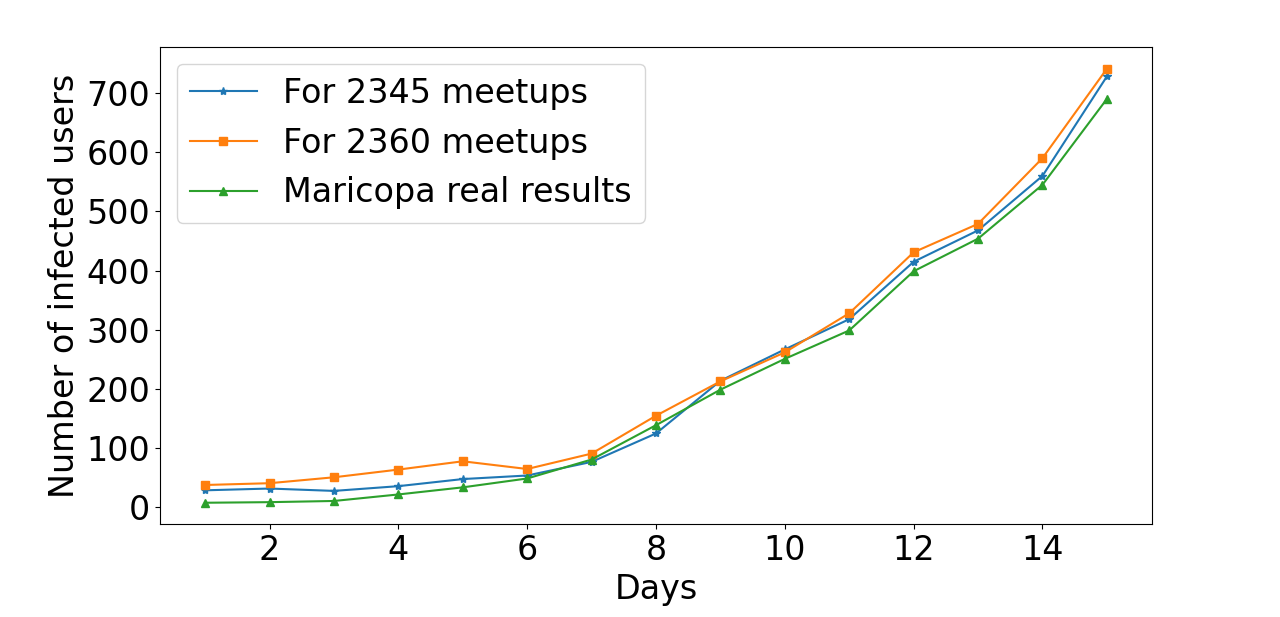}
	\caption{Case-6}
	\label{fig:case6}
\end{figure}

\textbf{Case-VI:} In this case, we work on Maricopa dataset with threshold probability of infection $\Theta = 0.9$, threshold time of contact $\tau_0 = 2$ minutes, threshold signal strength $\nu_0 = -0.55$ dBm and assuming number of meetups as 2345 and 2360. As given in Fig. \ref{fig:case6}, three curves are plotted, first is for 2345 meetups, the second is for 2360 meetups and the third is for real-world results of Maricopa.

\begin{figure}[h!]
\centering
	\includegraphics[height=4.5cm, width=8cm]{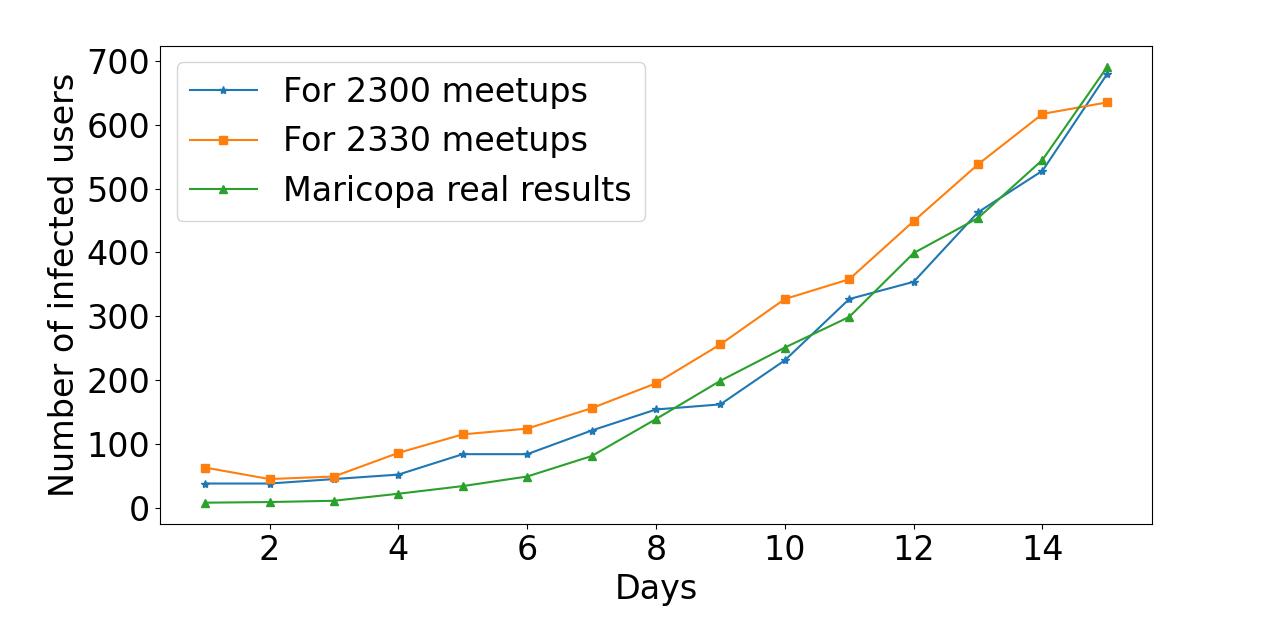}
	\caption{Case-7}
	\label{fig:case7}
\end{figure}

\textbf{Case-VII:} In this case, we work on Maricopa dataset with threshold probability of infection $\Theta = 0.9$, threshold time of contact $\tau_0 = 1$ minutes, threshold signal strength $\nu_0 = -0.50$ dBm and assuming number of meetups as 2300 and 2330. As given in Fig. \ref{fig:case7}, three curves are plotted, first is for 2300 meetups, the second is for 2330 meetups and the third is for real-world results of Maricopa.

\begin{figure}[h!]
\centering
	\includegraphics[height=4.5cm, width=8cm]{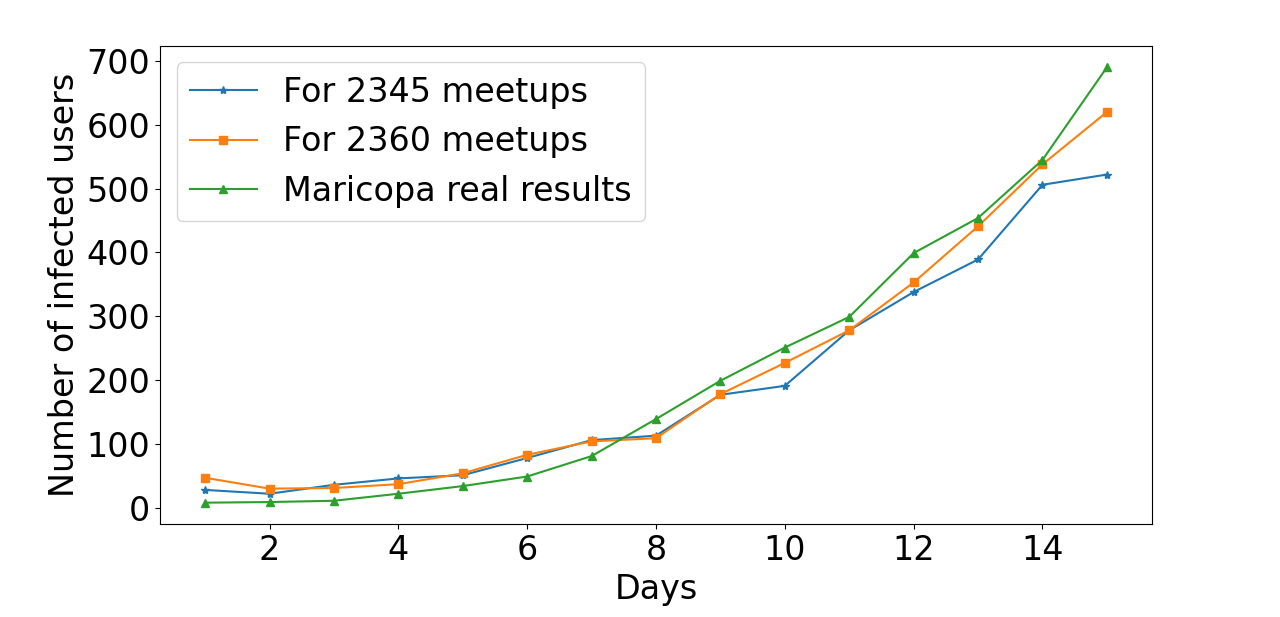}
	\caption{Case-8}
	\label{fig:case8}
\end{figure}

\textbf{Case-VIII:} In this case, we work on Maricopa dataset with threshold probability of infection $\Theta = 0.9$, threshold time of contact $\tau_0 = 1$ minutes, threshold signal strength $\nu_0 = -0.50$ dBm and assuming number of meetups as 2345 and 2360. As given in Fig. \ref{fig:case8}, three curves are plotted, first is for 2345 meetups, the second is for 2360 meetups and the third is for real-world results of Maricopa.

As we can see that the accuracy of the graphs is very good for some parameters like in case-1 and case-4 etc, but it doesn't mean that they always provide the right solution because meetups are directly dependent on people but in case, maybe with some government-imposed guidelines like lockdown, curfew can also be used to reduce the number of meetings between people. So, our model is used to track every user, so that the next chains can be broken and the spread of the virus can be reduced in the early stages only. It also takes care of each user individually by sending them their reports and all the needed precautions according to the risk levels of the individual user. And one more important thing to be noticed is that if at a certain day, according to the conditions of that day, future cases can also be calculated roughly just by taking the average number of meetups according to the condition, like in the condition of lockdown or curfew, the average number of meetups will be decreased and on a normal day, meetups can be more. So, with a rough idea of the current situation ( number of meetups nowadays), future prediction can also be done with our model. After having an average number of meetups, randomly two users are selected for each meetup and if any one of them is infected then with the help of algorithms, probabilities are calculated and finally, the rate of virus spread can be calculated and based on the results, actions can be taken like lockdown, curfew, vaccination, etc so that disastrous situations can be avoided. This also helps us get prepared for future conditions.

\section{Conclusion and Future Work}\label{ConFu}

In this work, we present a fog support framework, which can help individuals by keeping them safe and away from infected people and by advising them with the best precautions, guidelines, and medications. It also helps us predicting future conditions according to the present scenario of virus spread. In this model, we have used HMM and contact tracing via Bluetooth, and finally, implemented the framework with real-world data. From the real-world data analysis, we have concluded the effectiveness of the proposed model on different simulation settings for helping out taking precautionary steps in advance to stop the spread of the virus.

In the future, we'll propose a mechanism that can also take care of asymptomatic scenarios of COVID-19 infection. Moreover, we are also targeting to include the wider real-world data set from different countries to see the effectiveness of our proposed domain. Furthermore, the proposed model can also be extended to post real-time monitoring, prediction, and treatment of COVID-19 patients remotely with the involvement of doctors. To avoid the different drawbacks of Bluetooth, we also believe that the connectivity of the devices over cellular 5G technology \cite{9371426, 8783994} could further improve the transmission and the computation latencies in the networking model.        

\vspace{10pt}
\noindent \textbf{Acknowledgments:} The authors thank Mr. Tanay Jonwal and Mr. Lali Akhil Raj for their insightful discussions.




\begin{IEEEbiography}[{\includegraphics[width=1.1in,height=1.2in,clip,keepaspectratio]{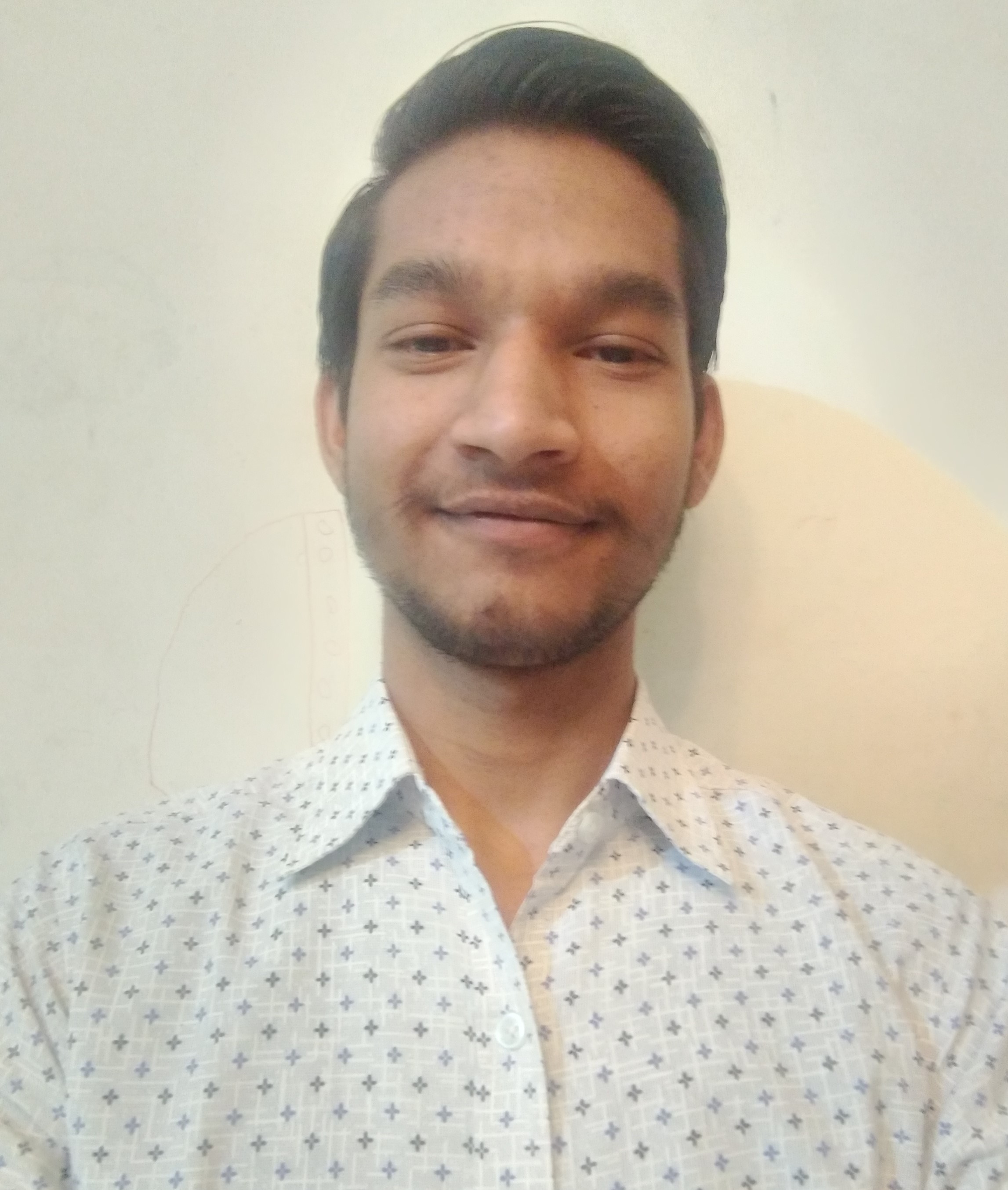}}]{Suraj Mahawar} is B.Tech 2nd year student in the Department of Computer Science and Engineering, Indian Institute of Technology (BHU) Varanasi, India. His interests include Algorithm Design and Mathematical modeling of real time problems.
\end{IEEEbiography}

\begin{IEEEbiography}[{\includegraphics[width=1.1in,height=1.2in,clip,keepaspectratio]{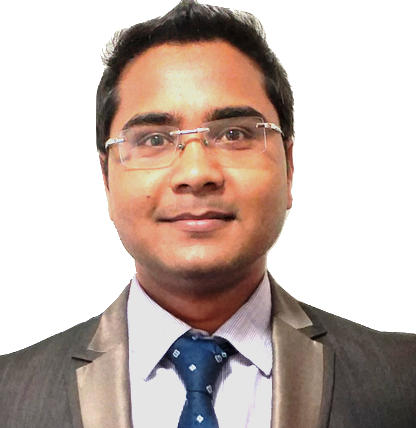}}]{Ajay Pratap} is an Assistant Professor with the Department of Computer Science and Engineering, Indian Institute of Technology (BHU) Varanasi, India. Before joining IIT (BHU), he was associated with the Department of Computer Science and Engineering, National Institute of Technology Karnataka (NITK) Surathkal, India, as an Assistant Professor from December 2019 to May 2020. He worked as a Postdoctoral Researcher in the Department of Computer Science at Missouri University of Science and Technology, USA, from August 2018 to December 2019. He completed his Ph.D. degree in Computer Science and Engineering from the Indian Institute of Technology Patna, India, in July 2018. His research interests include Cyber-Physical Systems, IoT-enabled Smart Environments, Mobile Computing and Networking, Statistical Learning, Algorithm Design for Next-generation Advanced Wireless Networks, Applied Graph Theory, and Game Theory. His current work is related to HetNet, Small Cells, Fog Computing, IoT, and D2D communication underlaying cellular 5G and beyond. His papers appeared in several international journals and conferences including IEEE Transactions on Mobile Computing, IEEE Transactions on Parallel and Distributed Systems, and IEEE LCN, etc. He has received several awards including the Best Paper Candidate Award and NSF travel grant for IEEE Smartcom'19 in the USA. 
\end{IEEEbiography}

\end{document}